\documentclass[onecolumn,floatfix,aps,nofootinbib, showpacs, showkeys]{revtex4}

\usepackage[dvips]{epsfig}
\usepackage[english]{babel}
\usepackage{bbm}
\usepackage{verbatim}
\usepackage{array}
\usepackage{amsmath}
\usepackage{multirow}
\usepackage{amsfonts}
\usepackage{amssymb}
\usepackage{hyperref}
\usepackage{leading}
\usepackage[dvipsnames]{xcolor}
\hypersetup{colorlinks=true,linkbordercolor=Blue,linkcolor=Blue, citecolor=Blue}
\usepackage{subeqnarray}
\usepackage{setspace}
\usepackage{placeins}
\usepackage{float}
\usepackage{indentfirst} 

\usepackage{gensymb}

\begin{document}

\title{Subliminal aspects concerning the Lounesto's classification}

\author{R. J. Bueno Rogerio$^{1}$} \email{rodolforogerio@unifei.edu.br}

\affiliation{$^{1}$Instituto de F\'isica e Qu\'imica, Universidade Federal de Itajub\'a - IFQ/UNIFEI, \\
Av. BPS 1303, CEP 37500-903, Itajub\'a - MG, Brazil.}


\begin{abstract}
{\textbf{Abstract.}} In the present communication we employ a split programme applied to spinors belonging to the regular and singular sectors of the Lounesto's classification \cite{lounestolivro}, looking towards unveil how it can be built (or defined) upon two spinors arrangement. We separate the spinors into two distinct parts and investigate to which class within the Lounesto's classification each part belong. The machinery here developed open up the possibility to better understand how spinors behave under such classification. As we shall see, the resulting spinor from the arrangement of other spinors (belonging to a distinct class or not) does not necessarily inherit the characteristics of the spinors that compose them, as example, such characteristics stands for the class, dynamic or the encoded physical information.
\end{abstract}


\maketitle

\section{Introduction}\label{intro}
Spinors are objects widely used in physics and also in mathematics. Firstly, their importance comes from the fact that they carry a rich information about the space-time where they are built, besides they play the central role to describing fermionic matter fields. Such a mathematical objects were firstly defined by \'{E}lie Cartan \cite{cartanbook} where he provided the following definition to a spinor ``\emph{A spinor is thus a sort of ``directed'' or ``polarised'' isotropic vector; a rotation about an axis through
an angle $2\pi$ changes the polarisation of this isotropic vector}''. Such entities may be defined without reference to the theory of representations of groups \cite{carmeli}. Spinors can be used without reference to relativity, but they arise naturally in discussions of
the Lorentz group. 

Due the importance of spinors, in a foundational work, Lounesto had the very idea to classify them \cite{lounestolivro}. The principle that he followed was an extensive and tedious algebraic analysis relating to spinors due to its bilinear covariants (physical information). Given this mathematical procedure, Lounesto assures the existence of only six classes of spinors, from which he denominated a group of three classes called ``Dirac spinors of the electron'' and the other remaining three spinors on a group denominated as ``Singular spinors with a light-like pole'' or only ``Singular spinors''. Having said that, the Lounesto's classification is taken as an algebraic classification. Several studies related to spinor and Lounesto's classification have been developed in recent times \cite[and references therein]{bonorapandora,beyondlounesto,rodolfoconstraints,revealing,cavalcanticlassification,cavalcanti4}.

The idea behind the present essay is based on the recent development \cite{rodolfoconstraints}, where it is shown a systematization/categorization on how to ascertain the spinor class due to its phases factor, without the necessity to evaluate all the bilinear forms. Here our focus is look towards analyse how spinors may be constituted. The mechanism that we use comes from the assumption that a spinor can be constructed from a arrangement of other spinors. In this vein, we investigate the conditions or the constraints that allow a spinor composition, also we analyse what kind of combinations are possible and what is the expected result. As we shall see, such a mechanism shows the impossibility of a generalization for spinor construction. Some classes are built given some algebraic constraints and other classes, as class 6, for example, do not guarantee that this procedure will be applied. As highlighted in \cite{rodolfoconstraints,cavalcanticlassification}, class 6 hold a very special case of spinors.

The paper is organized as it follows: in the next section we show the spinors separation method to be used along the article. In Sect.\ref{protocolregular} we take advantage of the method and investigate how spinors belonging to the regular sector of the Lounesto's classification are composed. Thus, in Sect.\ref{protocolsingular} we perform the very same analyse but now for singular spinors. Finally, in Sect.\ref{remarks}, we conclude. 

\section{Defining the spinorial detachment programme}\label{protocol}
The protocol that we will define, look towards searching the physical information encoded on the spinors. For this, we decompose a spinor as the sum of two (distinct) spinors. We want to show that: behind this mechanism there is a general rule --- or a composition law --- for spinors to belong to a certain class and, thus, only some possible combinations of spinors are allowed. Contrary to what can be imagined, we show that not every combination of spinors is valid or mathematically possible. Besides, it is possible to show that the physical information related to two spinors is not necessarily carried by the resulting spinor.

The programme to be employed here, is based on a spinor ``division''. Here we impose to the phases factors the following requirement $\alpha,\beta\in\mathbb{C}$. The last mentioned feature allow one to write the following $\alpha = Re\;\alpha + i{\rm Im}\;\alpha$ and $\beta = Re\;\beta + i{\rm Im}\;\beta$. 
Now, suppose the following spinor split
\begin{equation}
\psi_j = \left( \begin{array}{c}
Re\alpha \;\; \phi_R \\ 
Re\beta \;\;\phi_L
\end{array} \right) +  i \left( \begin{array}{c}
{\rm Im}\alpha \;\;\phi_R \\ 
{\rm Im}\beta \;\;\phi_L
\end{array} \right),
\end{equation}
in other words, it may be expressed as it follows
\begin{equation}\label{spinordecomposed}
\psi_j =\psi_k + \psi_l,
\end{equation}
note that we now have two distinct spinors, where the label $j, k$ and $l$ stands for the corresponding Lounesto's classes of each spinor and it runs $j,k,l=1,\ldots,6$. For the purpose of simplifying the notation, we omitted the spinor's momentum $(\boldsymbol{p})$. 

The relation presented in \eqref{spinordecomposed} is the protocol to \emph{separate} a spinor. Now, we have a spinor carrying the real part of the phases ($Re$) and another spinor carrying the imaginary part of the phases (${\rm Im}$). What we shall check here is whether the combination ($+$) of spinors preserves the class and what is the outcome when one sum different spinor classes. Let $\Gamma$ be a set of bilinear forms of a given spinor \cite{crawford1,crawford2}
\begin{equation}
\Gamma^{j} = \{\sigma, \omega, \boldsymbol{J}, \boldsymbol{K}, \boldsymbol{S}\},
\end{equation}
where $j$ stands for the spinor class. When performing the procedure defined in \eqref{spinordecomposed}, automatically a superposition of the spinor's physical information is observed --- thus, such a feature translate into 
\begin{equation}\label{gamasomado}
\Gamma^{j} = \Gamma^{k}\cup\Gamma^{l}.
\end{equation}
Note that \eqref{gamasomado} tell us that a given set of bilinear amounts can be obtained from an union of two distinct (or even equal) sets of bilinear amounts. What should be clear to the reader is that the above definition is not exactly a summation of each one of the bilinear forms separately but it stands for a combination of bilinear forms that, when added together, provide a new set of bilinear forms. 

\subsection{On the regular spinors framework}\label{protocolregular}
A single-helicity spinor in the rest-frame referential is defined as follows \cite{interplay,rodolfoconstraints} 
 \begin{eqnarray}\label{psigeral}
\psi(k^{\mu}) = \left(\begin{array}{c}
\alpha\phi_{R}^{+}(k^{\mu})\\
\beta\phi_{L}^{+}(k^{\mu})
\end{array}
\right), \;\;\mbox{and} \;\; \psi(k^{\mu}) = \left(\begin{array}{c}
\alpha\phi_{R}^{-}(k^{\mu})\\
\beta\phi_{L}^{-}(k^{\mu})
\end{array}
\right),
\end{eqnarray}
in which we have defined the $k^{\mu}$ rest-frame momentum as 
\begin{equation}
k^{\mu}\stackrel{def}{=}\bigg(m,\; \lim_{p\rightarrow 0}\frac{\boldsymbol{p}}{p}\bigg), \; p=|\boldsymbol{p}|.
\end{equation}
Commonly, the spinorial components in the rest-frame referential reads
\begin{equation}
\phi_{R/L}^{+}(k^{\mu})  = \sqrt{m}\left(\begin{array}{c}
\cos(\theta/2)e^{-i\phi/2} \\ 
\sin(\theta/2)e^{i\phi/2}
\end{array} \right), 
\end{equation}
and 
\begin{equation}
\phi_{R/L}^{-}(k^{\mu}) = \sqrt{m}\left(\begin{array}{c}
-\sin(\theta/2)e^{-i\phi/2} \\ 
\cos(\theta/2)e^{i\phi/2}
\end{array} \right), 
\end{equation}
where the phases factors $\alpha$ and $\beta$ $\in\mathbbm{C}$ and the only requirement under the such factors, comes from the orthonormal relation, it stands for the regular spinor's case $\alpha\beta^*+\alpha^*\beta\propto m$, where $m$ stands for the mass of a particle.
The upper indexes $\pm$ refers to the corresponding helicity of each component, for more details the Reader is cautioned to check \cite{mdobook}. 

Note that if one wish to define such rest spinors in an momentum arbitrary referential, such task is accomplished under action of the Lorentz boosts operator, which reads
\begin{eqnarray}\label{boostoperator}
e^{i\kappa.\varphi} = \sqrt{\frac{E+m}{2m}}\left(\begin{array}{cc}
\mathbbm{1}+\frac{\vec{\sigma}.\hat{p}}{E+m} & 0 \\ 
0 & \mathbbm{1}-\frac{\vec{\sigma}.\hat{p}}{E+m}
\end{array} \right),
\end{eqnarray}
as usually defined $\cosh\varphi = E/m, \sinh\varphi=p/m$, and  $\hat{\boldsymbol{\varphi}} = \hat{\boldsymbol{p}}$, yielding the following relation 
\begin{equation}
\psi(p^{\mu}) = e^{i\kappa.\varphi}\psi(k^{\mu}).
\end{equation}

Now, consider the Dirac operator acting on \eqref{psigeral}
\begin{equation}\label{diraceq}
(\gamma_{\mu}p^{\mu} - m)\psi(\boldsymbol{p})=0,
\end{equation}
the Dirac operator acting over the $\psi(\boldsymbol{p})$ spinor provides
\begin{eqnarray}\label{diraceq2}
\gamma_{\mu}p^{\mu}\psi(\boldsymbol{p}) = \Bigg[E\left(\begin{array}{cc}
0 & \mathbbm{1} \\ 
\mathbbm{1} & 0
\end{array} \right)+ p\left(\begin{array}{cc}
0 & \vec{\sigma}\cdot\hat{p} \\ 
-\vec{\sigma}\cdot\hat{p} & 0
\end{array} \right)\Bigg]\left(\begin{array}{c}
\alpha\phi_R(\boldsymbol{p})\\
\beta\phi_L(\boldsymbol{p})
\end{array}
\right),
\end{eqnarray}
where the operator $\vec{\sigma}\cdot\hat{p}$ stands for the helicity operator, where $\sigma$ stands for the Pauli matrices and $\hat{p}$ is the unit momentum vector. In order to proceed with the calculations, we take into account the spinors which carry positive helicity in \eqref{psigeral} and, then, we obtain the following relation\footnote{Some mathematical steps were omitted from this quick derivation, due to recurrence this appears in the literature. However, some details should be highlighted, to right ascertain the Eq.\eqref{diraceq3} it should be keep in mind that $\phi_{R}(\boldsymbol{0}) =\pm\phi_{L}(\boldsymbol{0})$, where the upper (lower) index stands for particle (antiparticle) case \cite{ryder,gaioliryder,ahluwalia1ryder}. }
\begin{eqnarray}\label{diraceq3}
\gamma_{\mu}p^{\mu}\psi(\boldsymbol{p}) = m\left(\begin{array}{c}
\beta\phi_R^{+}(\boldsymbol{p})\\
\alpha\phi_L^{+}(\boldsymbol{p})
\end{array}
\right),
\end{eqnarray}
up to our knowledge, \eqref{diraceq} is only fulfilled if $\alpha = \beta$, otherwise the Dirac dynamic is not reached. The last result combined with Table 1 in \cite{rodolfoconstraints} lead to the observation that only spinors belonging to class 2 within Lounesto's classification, under the requirement $\alpha=\beta$, satisfy the Dirac equation. We remark that dynamics is maintained when combining class 2  spinors with $\alpha=\beta$; in other words, the sum of two spinors that satisfy Dirac's dynamic necessarily provide a spinor that satisfies Dirac's equation --- holding the Dirac equation linearity. 

We emphasize that the Lounesto's classification is geometric \cite{lounestolivro}, that is, it is based on the spinor bilinear forms (physical observable) and a strong link between such quantities, namely Fierz-Pauli-Kofink identities. Therefore, such classification does not refer to dynamics. Lounesto, based on Crawford's bispinoral densities derived in \cite[and references therein]{crawford1}, when he derives the 16 bilinear forms for spinors that himself calls ``Dirac spinors'', develops the analysis taking into account an arbitrary spinor that supposedly satisfies Dirac's dynamics. However, Lounesto does not verify (explicitly) if all the 3 classes of Dirac spinors satisfy the Dirac's equation, moreover, he does not even mention whether this is a necessary condition for Dirac spinors to belong to these classes. We emphasize here, is that the result we found in equation \eqref{diraceq3} is important, as it shows a strong condition in which only one class of spinors of the so-called `` Dirac spinors '' satisfies the Dirac dynamics. Not necessarily every spinor that is classified as a Dirac spinor, within the Lounesto classification, must obey Dirac's dynamics.
\newpage
Accordingly to the Table 1 of \cite{rodolfoconstraints} we may perform the following analysis:
\begin{description}
\item[1)] \underline{$\alpha, \beta \in \mathbb{C}$ with $\alpha\neq\beta$ (class 1)}: 
\end{description}
In view of the protocol introduced above, we start analysing the first case, which lead to
\begin{eqnarray}
\psi_1 =  \left( \begin{array}{c}
Re\alpha \;\; \phi_R \\ 
Re\beta \;\;\phi_L
\end{array} \right) + i \left( \begin{array}{c}
{\rm Im}\alpha \;\;\phi_R \\ 
{\rm Im}\beta \;\;\phi_L
\end{array} \right),
\end{eqnarray}
thus, such mechanism brings to the light the following relation
\begin{equation}
\psi_1 = \psi_2+\psi_2.
\end{equation}
Note that a class 1 spinor may be built upon two spinors belonging to class 2. 
However, although class 1 spinors may be composed by class 2 spinors, they do not satisfy the Dirac's dynamic. 

\begin{description}
\item[2)] \underline{$\alpha \in\mathbb{C}$ and  $\beta \in {\rm I\!R}  $ (class 1)}: 
\end{description}
Now, note that
\begin{eqnarray}
\psi_1 =  \left( \begin{array}{c}
Re\alpha \;\; \phi_R \\ 
Re\beta \;\;\phi_L
\end{array} \right) + i \left( \begin{array}{c}
{\rm Im}\alpha \;\;\phi_R \\ 
0
\end{array} \right),
\end{eqnarray}
leading to
\begin{equation}
\psi_1 = \psi_2+\psi_6,
\end{equation}
and, thus, we remark a new possibility to write a spinor which belong to class 1.
\begin{description}
\item[3)] \underline{$\alpha \in\mathbb{C}$ and  $\beta \in {\rm Im}  $ (class 1)}: 
\end{description}
For this case at hands we have 
\begin{eqnarray}
\psi_1 =  \left( \begin{array}{c}
Re\alpha \;\; \phi_R \\ 
0
\end{array} \right) + i \left( \begin{array}{c}
{\rm Im}\alpha \;\;\phi_R \\ 
{\rm Im}\beta \;\;\phi_L
\end{array} \right),
\end{eqnarray}
and the only possibility stands for
\begin{equation}
\psi_1 = \psi_6+\psi_2.
\end{equation}

\begin{description}
\item[4)] \underline{$\alpha, \beta \in \mathbb{C}$  with $\alpha=\beta$ (class 2)}:
\end{description}
Such constraints leads to 
\begin{eqnarray}
\psi_2 =  \left( \begin{array}{c}
Re\alpha \;\; \phi_R \\ 
Re\alpha \;\;\phi_L
\end{array} \right) + i \left( \begin{array}{c}
{\rm Im}\alpha \;\;\phi_R \\ 
{\rm Im}\alpha \;\;\phi_L
\end{array} \right),
\end{eqnarray}
the above calculations allow one to write
\begin{equation}
\psi_2 = \psi_2+\psi_2.
\end{equation}
Given the less restrictive requirements for a spinor to belong to class 2, meantime, this is the only possibility to write them as a combination of other spinors.

\begin{description}
\item[5)] \underline{$\alpha \in{\rm Im}$ and  $\beta \in {\rm I\!R}  $ (class 3)}: 
\end{description}
Here we find two quite peculiar situations, the above requirements provides the following relation
\begin{eqnarray}
\psi_3 =  \left( \begin{array}{c}
0 \\ 
Re\beta \;\;\phi_L
\end{array} \right) + i \left( \begin{array}{c}
{\rm Im}\alpha \;\;\phi_R \\ 
0
\end{array} \right),
\end{eqnarray}
which translates into
\begin{equation}
\psi_3 = \psi_6+\psi_6.
\end{equation}

\begin{description}
\item[6)] \underline{$\alpha \in {\rm I\!R} $ and  $\beta \in {\rm Im}  $ (class 3)}: 
\end{description}
It allows one to define
\begin{eqnarray}
\psi_3 =  \left( \begin{array}{c}
Re\alpha \;\; \phi_R \\ 
0
\end{array} \right) + i \left( \begin{array}{c}
0 \\ 
{\rm Im}\beta \;\;\phi_L
\end{array} \right),
\end{eqnarray}
leading to
\begin{equation}
\psi_3 = \psi_6+\psi_6.
\end{equation}
Interestingly enough, class 3 spinors can only be defined as a combination of two class 6 spinors.

Thus, the above results can be summarized as it follows:
\begin{table}[H]
\centering
\begin{tabular}{ccc}
\hline
\:\: Class\:\: & \:\: Spinorial combination\:\: & \:\: Phases constraints\:\: \\ 
\hline 
\hline 
$1$ & $\psi_2+ \psi_2$ & $\forall\alpha,\beta\in {\rm I\!R}$ or $\forall\alpha,\beta\in {\rm Im}$ 
\vspace{0.1cm}\\
$1$ & $\psi_2+ \psi_6$ & $\forall\alpha\in\mathbb{C}$ and $\forall\beta\in {\rm I\!R}$ or $\forall\alpha\in\mathbb{C}$ and $\forall\beta\in {\rm Im}$ 
\vspace{0.1cm}\\ 
$2$ & $\psi_2+ \psi_2$ & $\alpha,\beta\in\mathbb{C} | \alpha=\beta$
\vspace{0.1cm}\\
$3$ & $\psi_6+ \psi_6$ & $\forall\alpha\in{\rm Im}$ and $\forall\beta\in {\rm I\!R}$ or $\forall\alpha\in{\rm Im}$ and $\forall\beta\in {\rm Im}$  
\\
\hline 
\end{tabular} 
\label{tabela1}
\caption{Spinorial combination for the Lounesto's regular sector}
\end{table}
\noindent We highlight that to obtain certain classes, we face some restrictions, e.g., the impossibility to construct a spinor belonging to class 6, it does not admit to be written as a combination of two distinct spinors.

\subsection{On the singular spinors framework}\label{protocolsingular}
In this section, we look towards applying the previous algorithm on dual-helicity spinors. Dual-helicity spinors can be defined as \cite{mdobook, interplay, rodolfoconstraints,cavalcanticlassification}
 \begin{eqnarray}\label{dualhelicity1}
\psi = \left(\begin{array}{c}
\alpha\Theta\phi_L^{* \pm}\\
\beta\phi_L^{\pm}
\end{array}
\right), 
\end{eqnarray}
where $\Theta$ it the well-known Wigner Time-reversal operator 
\begin{equation}
\Theta = \left(\begin{array}{cc}
0 & -1 \\ 
1 & \;\;0
\end{array} \right).
\end{equation}
Taking advantage of Table 2 presented in \cite{rodolfoconstraints}, and the spinor defined in \eqref{dualhelicity1}, one is able to define the following
\begin{description}
\item[1)] \underline{$\alpha ,\beta \in\mathbb{C}$ with $|\alpha|^2\neq|\beta|^2$ (class 4)}: 
\end{description}
For this case we have
\begin{eqnarray}
\psi_4 =  \left( \begin{array}{c}
Re\alpha \;\; \Theta\phi_L^{*} \\ 
Re\beta \;\;\phi_L
\end{array} \right) + i \left( \begin{array}{c}
{\rm Im}\alpha \;\;\Theta\phi_L^{*} \\ 
{\rm Im}\beta \;\;\phi_L
\end{array} \right),
\end{eqnarray}
which provide the following relations
\begin{eqnarray}
&&\psi_4 = \psi_4+ \psi_4, 
\\
&&\psi_4 = \psi_5+ \psi_5, 
\\
&&\psi_4 = \psi_4+ \psi_5. 
\end{eqnarray}
showing a wide variety of combinations.
\begin{description}
\item[2)] \underline{$\alpha ,\beta \in\mathbb{C}$ with $|\alpha|^2=|\beta|^2$ (class 5)}: 
\end{description}
Furnishing the following detachment
\begin{eqnarray}
\psi_5 =  \left( \begin{array}{c}
Re\alpha \;\; \Theta\phi_L^{*} \\ 
Re\beta \;\;\phi_L
\end{array} \right) + i \left( \begin{array}{c}
{\rm Im}\alpha \;\;\Theta\phi_L^{*} \\ 
{\rm Im}\beta \;\;\phi_L
\end{array} \right),
\end{eqnarray}
which can be divided into 
\begin{eqnarray}
&&\psi_5 = \psi_4+ \psi_4, 
\\
&&\psi_5 = \psi_5+ \psi_5. 
\end{eqnarray}
Note that the Majorana spinor (which describes the neutrino) can be written in terms of two non-neutral spinors. 
\begin{description}
\item[3)] \underline{$\alpha \in \mathbb{C}$ and $\beta\in {\rm I\!R}$ with $|\alpha|^2\neq|\beta|^2$ (class 4)}: 
\end{description}
Note that
\begin{eqnarray}
\psi_4 =  \left( \begin{array}{c}
Re\alpha \;\; \Theta\phi_L^{*} \\ 
Re\beta \;\;\phi_L
\end{array} \right) + i \left( \begin{array}{c}
{\rm Im}\alpha \;\;\Theta\phi_L^{*} \\ 
0
\end{array} \right),
\end{eqnarray}
yielding  the following possibilities
\begin{eqnarray}
&&\psi_4 = \psi_4+ \psi_6, 
\\
&&\psi_4 = \psi_5+ \psi_6. 
\end{eqnarray}

\begin{description}
\item[4)] \underline{$\alpha \in\mathbb{C}$ and $\beta\in {\rm I\!R}$ with $|\alpha|^2=|\beta|^2$ (class 5)}: 
\end{description}
Such conditions above allow one to write
\begin{eqnarray}
\psi_5 =  \left( \begin{array}{c}
Re\alpha \;\; \Theta\phi_L^{*} \\ 
Re\beta \;\;\phi_L
\end{array} \right) + i \left( \begin{array}{c}
{\rm Im}\alpha \;\;\Theta\phi_L^{*} \\ 
0
\end{array} \right),
\end{eqnarray}
resulting in 
\begin{eqnarray}
\psi_5 = \psi_4+ \psi_4. 
\end{eqnarray}

\begin{description}
\item[5)] \underline{$\alpha\in\mathbb{C}$ and $\beta \in{\rm Im} $ with $|\alpha|^2\neq|\beta|^2$ (class 4)}: 
\end{description}
Now, notice
\begin{eqnarray}
\psi_4 =  \left( \begin{array}{c}
Re\alpha \;\; \Theta\phi_L^{*} \\ 
0
\end{array} \right) + i \left( \begin{array}{c}
{\rm Im}\alpha \;\;\Theta\phi_L^{*} \\ 
{\rm Im}\beta \;\;\phi_L
\end{array} \right),
\end{eqnarray}
furnishing
\begin{eqnarray}
&&\psi_4 = \psi_4+ \psi_6, 
\\
&&\psi_4 = \psi_5+ \psi_6. 
\end{eqnarray}

\begin{description}
\item[6)] \underline{$\alpha\in\mathbb{C}$ and $\beta \in{\rm Im} $ with $|\alpha|^2=|\beta|^2$ (class 5)}: 
\end{description}
Thus,
\begin{eqnarray}
\psi_5 =  \left( \begin{array}{c}
Re\alpha \;\; \Theta\phi_L^{*} \\ 
0
\end{array} \right) + i \left( \begin{array}{c}
{\rm Im}\alpha \;\;\Theta\phi_L^{*} \\ 
{\rm Im}\beta \;\;\phi_L
\end{array} \right),
\end{eqnarray}
making explicit the relation
\begin{eqnarray}
\psi_5 = \psi_4+ \psi_6. 
\end{eqnarray}

\begin{description}
\item[7)] \underline{$\alpha\in{\rm Im}$ and $\beta \in {\rm I\!R} $ with $|\alpha|^2\neq|\beta|^2$ (class 4)}: 
\end{description}
Consequently,
\begin{eqnarray}
\psi_4 =  \left( \begin{array}{c}
0 \\ 
Re\beta \;\;\phi_L
\end{array} \right) + i \left( \begin{array}{c}
{\rm Im}\alpha \;\;\Theta\phi_L^{*} \\ 
0
\end{array} \right),
\end{eqnarray}
yielding the unique relation
\begin{eqnarray}
\psi_4 = \psi_6+ \psi_6. 
\end{eqnarray}

\begin{description}
\item[8)] \underline{$\alpha\in{\rm Im}$ and $\beta \in {\rm I\!R} $ with $|\alpha|^2=|\beta|^2$ (class 5)}: 
\end{description}
And finally we have
\begin{eqnarray}
\psi_5 =  \left( \begin{array}{c}
0 \\ 
Re\beta \;\;\phi_L
\end{array} \right) + i \left( \begin{array}{c}
{\rm Im}\alpha \;\;\Theta\phi_L^{*} \\ 
0
\end{array} \right),
\end{eqnarray}
which provide
\begin{eqnarray}
\psi_5 = \psi_6+ \psi_6. 
\end{eqnarray}
Where two Weyl spinors (massless neutrino) together compose a Majorana's neutrino.
 
In general grounds we may display the above results as
\begin{table}[H]
\centering
\begin{tabular}{ccc}
\hline
\:\: Class\:\: & \:\: Spinorial combination\:\: & \:\: Phases constraints\:\: \\ 
\hline 
\hline 
$4$ & $\psi_4+ \psi_4$ & 
\vspace{0.1cm}\\
$4$ & $\psi_5+ \psi_5$ & $\alpha ,\beta \in\mathbb{C}$ with $|\alpha|^2\neq|\beta|^2$ 
\vspace{0.1cm}\\ 
$4$ & $\psi_4+ \psi_5$ & 
\vspace{0.1cm}\\\\ 
$4$ & $\psi_4+ \psi_6$ & $\alpha \in \mathbb{C}$ and $\beta\in {\rm I\!R}$ with $|\alpha|^2\neq|\beta|^2$ or $\alpha \in \mathbb{C}$ and $\beta\in {\rm Im}$ with $|\alpha|^2\neq|\beta|^2$   
\vspace{0.1cm}\\
$4$ & $\psi_5+ \psi_6$ & $\alpha \in \mathbb{C}$ and $\beta\in {\rm I\!R}$ with $|\alpha|^2\neq|\beta|^2$ or $\alpha \in \mathbb{C}$ and $\beta\in {\rm Im}$ with $|\alpha|^2\neq|\beta|^2$ 
\vspace{0.1cm}\\\\ 
$4$ & $\psi_6+ \psi_6$ & $\alpha\in{\rm Im}$ and $\beta \in {\rm I\!R} $ with $|\alpha|^2\neq|\beta|^2$ 
\vspace{0.1cm}\\\\ 
$5$ & $\psi_4+ \psi_4$ &  $\alpha ,\beta \in\mathbb{C}$ with $|\alpha|^2\neq|\beta|^2$ or $\alpha\in\mathbb{C}$ and $\beta \in {\rm I\!R} $ with $|\alpha|^2\neq|\beta|^2$
\vspace{0.1cm}\\ 
$5$ & $\psi_5+ \psi_5$ & $\alpha ,\beta \in\mathbb{C}$ with $|\alpha|^2\neq|\beta|^2$ 
\vspace{0.1cm}\\\\ 
$5$ & $\psi_4+ \psi_6$ & $\alpha \in \mathbb{C}$ and $\beta\in {\rm Im}$ with $|\alpha|^2=|\beta|^2$    
\vspace{0.1cm}\\ 
$5$ & $\psi_6+ \psi_6$ & $\alpha\in{\rm Im}$ and $\beta \in {\rm I\!R} $ with $|\alpha|^2=|\beta|^2$ 
\vspace{0.1cm}\\
\hline 
\end{tabular} 
\label{tabela2}
\caption{Spinorial combination for the Lounesto's singular sector}
\end{table}

\section{Final Remarks}\label{remarks}

In the present communication we delved into an investigation searching for complementary information about how spinors may be constituted/constructed from an arrangement between other spinors within Lounesto’s classification. As one can see, spinors can be written as a combination of other distinct spinors. Nonetheless, it does not hold true for all regular spinors classes, in which some specific classes must be combined to lead to a certain resulting class, as the case of classes 2 and 3, where only very restricted combinations are valid to define them, as it can be seen in Table I. Note that the same do not hold true for the singular spinors, which allow a range of possibilities of construction, check for Table II. Such a procedure unveils that it is possible to cover all of the Lounesto’s classes except class 6, which do not allow to be written as a combination of any other spinor.

Moreover, driven by the programme developed here, it is easy to see that when the spinor detach protocol is applied, the physical information do not necessarily is carried through the resulting spinor, as example, it does not hold the class, dynamic or even physical information. 

Interesting enough, from an inspection of Table II, when dealing with the singular sector of the Lounesto's classification, we may construct \emph{neutral} spinors from a combination of \emph{non-neutral} spinors, as the case presented in the rows 7, 9 and 10. Notice that the same akin reasoning can be extended for the case of class 4 spinors, which can be built upon two neutral spinors, as the case in the rows 2 and 5. We emphasize that a similar connection between both Lounesto's sections, as shown in \cite{rodolfoconstraints}, can be performed here, however, no relevant physical information is disclosed.

\section{Acknowledgements}
RJBR thanks CNPq Grant N$^{\circ}$. 155675/2018-4 for the financial support.

\bibliographystyle{unsrt}
\bibliography{refs}

\end{document}